\documentclass[american,english,preprint,amsmath,amssymb,aps,pre]{revtex4-1}
\usepackage[T1]{fontenc}
\usepackage[latin9]{inputenc}
\usepackage{color}
\usepackage{amsmath}
\usepackage{amssymb}
\usepackage{stmaryrd}
\usepackage{graphicx}
\usepackage{esint}

\makeatletter
\@ifundefined{textcolor}{}
{%
 \definecolor{BLACK}{gray}{0}
 \definecolor{WHITE}{gray}{1}
 \definecolor{RED}{rgb}{1,0,0}
 \definecolor{GREEN}{rgb}{0,1,0}
 \definecolor{BLUE}{rgb}{0,0,1}
 \definecolor{CYAN}{cmyk}{1,0,0,0}
 \definecolor{MAGENTA}{cmyk}{0,1,0,0}
 \definecolor{YELLOW}{cmyk}{0,0,1,0}
}

\usepackage[english]{babel}
\usepackage{subfigure}

\usepackage{babel}

\makeatother

\usepackage{babel}
\begin{document}

\title{Random Rectangular Graphs}

\author{Ernesto Estrada}

\author{Matthew \foreignlanguage{american}{Sheerin}}

\address{Department of Mathematics and Statistics, University of Strathclyde,
26 Richmond Street, Glasgow, G1 1XH, U.K. }

\maketitle
A generalization of the random geometric graph (RGG) model is proposed
by considering a set of points uniformly and independently distributed
on a rectangle of unit area instead of on a unit square $\left[0,1\right]^{2}.$
The topological properties of the \textit{\textcolor{black}{random
rectangular graphs}} (RRGs) generated by this model are then studied
as a function of the rectangle sides lengths $a$ and $b=1/a$, and
the radius $r$ used to connect the nodes. When $a=1$ we recover
the RGG, and when $a\rightarrow\infty$ the very elongated rectangle
generated resembles a one-dimensional RGG. We obtain here analytical
expressions for the average degree, degree distribution, connectivity,
average path length and clustering coefficient for RRG. These results
provide evidence that show that most of these properties depend on
the connection radius and the side length of the rectangle, usually
in a monotonic way. The clustering coefficient, however, increases
when the square is transformed into a slightly elongated rectangle,
and after this maximum it decays with the increase of the elongation
of the rectangle. We support all our findings by computational simulations
that show the goodness of the theoretical models proposed for RRGs.
\medskip{}


\section{Introduction}

The use of graphs for representing physical systems is becoming ubiquitous
in many areas of theoretical and applied physics \cite{Michaelbook}.
We can mention the use of graphs in statistical mechanics and condensed
matter physics, for solving Feynman integrals as well as in the study
of quantum phenomena \cite{Michaelbook,Quantum Graphs}. More recently,
the use of graphs has been very broadened by their application in
the analysis of complex systems \cite{EstradaBook,LucianoReview,NewmanReview}.
In this case, those graphs receive the name of complex networks, due
to the fact that they represent the skeleton of complex interconnected
systems. In this case, networks are used to study a variety of physical
scenarios, ranging from social and infrastructural, to biological
and ecological ones. Here, we will use the terms graphs and networks
interchangeably. When graphs are used to represent real-world physical
systems it is necessary to have at our disposal some null model that
allows us to evaluate which properties of the system have arisen from
their connectivity pattern. In this sense, the common election is
the use of random graphs. These are graphs with the same number of
nodes and edges as the one under study, but in which the connection
between the nodes is made randomly and independently \cite{Random graphs in networks}.
There are several of these random models of great usability in current
network theory, such as the Erdös-Rényi \cite{Erdos-Renyi}, the Barabási-Albert
\cite{Barabasi-Albert} or the Watts-Strogatz \cite{Watts-Strogatz}
model to mention just three.

In many real-world scenarios the networks emerge under certain geometrical
constraints. This is the case of the so-called spatial networks \cite{Spatial networks},
which include infrastructural networks such as road networks, airport
transportation networks, etc., \cite{Spatial networks} and certain
biological networks such as brain networks or the networks representing
the proximity of cells in a biological tissue (see \cite{EstradaBook}).
The list also includes the networks of patches and corridors in a
landscape \cite{Landscape networks}, the networks of galleries in
animal nests \cite{Termites,Ants}, and the networks of fractures
in rocks \cite{Rock fractures}, among others. The classical election
of a random graph used to represent these systems are the so-called
random geometric graphs \cite{Penrose,Dall Christense}. Here the
term \textit{\textcolor{black}{random geometric graph}} (RGG) is reserved
for the case in which the nodes of the graph are distributed randomly
and independently in a unit square and two nodes are connected if
they are inside a disk of a given radius centered at one of the nodes.
Other graphs in which the edges are constructed by using different
geometric rules will be named here generically as \textit{\textcolor{black}{random
proximity graphs}}.

RGGs have found important applications in the area of wireless communication
devices \cite{RGG wireless,RGG communication,RGG comm}, such as mobile
phones, wireless computing systems, wireless sensor networks, etc.
This was indeed the first application in mind when Gilbert proposed
the very first RGG model \cite{Gilbert model}. RGGs have also found
applications in areas such as modeling of epidemic spreading in spatial
populations, which may include cases such the spreading of worms in
a computer network, viruses in a human population, or rumors in a
social network \cite{Spatial connectivity,RGG Sync,Worm Epidemics,RGG spreading,RPG epidemics}.
RGGs have been used to describe how cities have been evolving under
the geometric constraints imposed by their geographic locations \cite{RPG cities}.
For a wider perspective on the applications of spatial graphs the
reader is referred to the review \cite{Spatial networks}.

In all the previously mentioned real-world scenarios, the shape of
the location in which the nodes of the graph are distributed may play
a fundamental role in the topological and dynamical properties of
the resulting graphs. That is, it is intuitive to think that the connectivity,
distance, clustering and other fundamental topological properties
of the graphs are affected if we, for instance, elongate the unit
square in which the points are distributed. Here, we develop a new
model that generalizes the RGG by allowing the embedding of the nodes
in a unit rectangle instead of a unit square. Our main goal is to
investigate how the elongation of a unit square influences the topological
properties of the graphs generated by the model. These generalized
graphs will be named here the \textit{random rectangular graphs} (RRGs).
In this work we concentrate on the influence of the length of the
rectangle on the topological properties of the graphs emerging on
them, such as their average degree, connectivity, degree distribution,
average path length and clustering coefficient. In particular, we
find analytical expressions and bounds for all of them and provide
computational evidence of the goodness of these approaches for relatively
large RRGs.

\section{Definition of the model}

The RGG is defined by distributing uniformly and independently $n$
points in the unit $d$-dimensional cube $[0,1]^{d}$ \cite{Penrose}.
Then, two points are connected by an edge if their (Euclidean) distance
is at most $r$, which is a given fixed number known as the connection
radius. 

\begin{figure}
\centering \subfigure[]{}\includegraphics[width=0.5\textwidth]{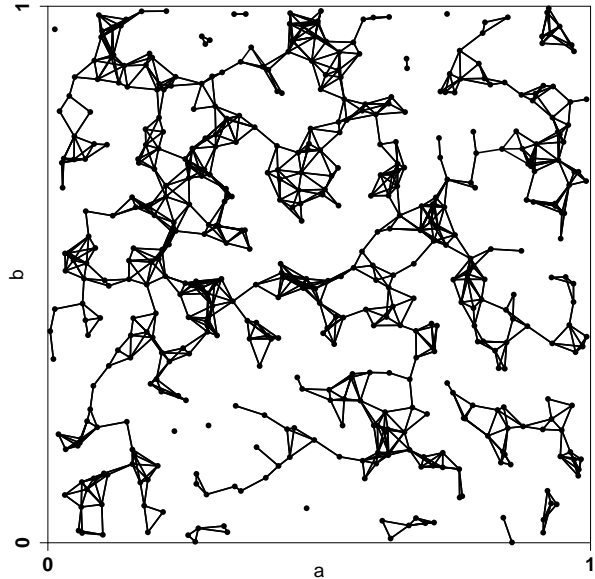}

\bigskip{}

\subfigure[]{}\includegraphics[width=1\textwidth]{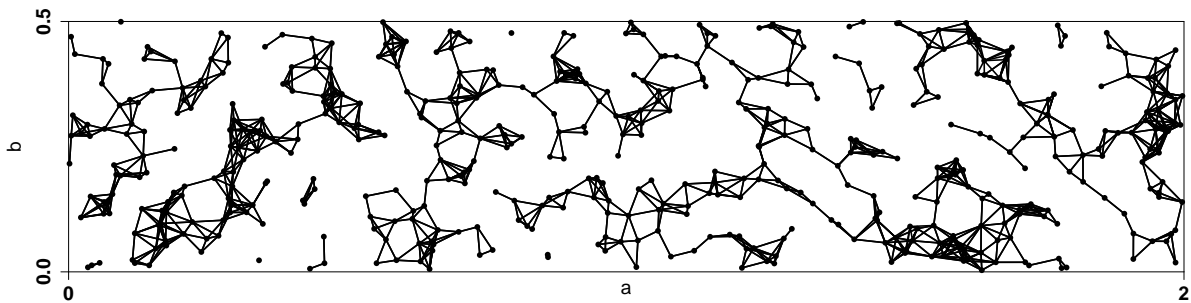}

\protect\protect\protect\caption{Illustration of two random rectangular graphs with $a=1$ (top), which
corresponds to a random geometric graph on a unit square and with
$a=2$ ($b=0.5$) (bottom). Both graphs are built with 500 nodes and
1750 edges.}

\label{RGG and RRG} 
\end{figure}

Let us now define a unit hyperrectangle as the Cartesian product $[a_{1},b_{1}]\times[a_{2},b_{2}]\times\cdots\times[a_{d},b_{d}]$
where $a_{i},b_{i}\in\mathbb{R},a_{i}\le b_{i}$, and $1\le i\le d$.
Hereafter we will restrict ourselves to the 2-dimensional case, which
corresponds to a rectangle of unit area, which we will call the unit
rectangle. Now, the RRG is defined by distributing uniformly and independently
$n$ points in the unit rectangle $[a,b]$ and then connecting two
points by an edge if their (Euclidean) distance is at most $r$. It
is evident that the only change we have introduced here is to consider
a rectangle of unit area instead of the analogous square. The rest
of the construction process remains the same as for the RGG. This
means that $RRG\rightarrow RGG$ as $\left(a/b\right)\rightarrow1$.
In this sense we can say that the RRG is a generalization of the RGG.
In Fig. \ref{RGG and RRG} we illustrate an RGG and an RRG constructed
with the same number of nodes and edges.

An interesting question is what happens at the other extreme, when
$a\rightarrow\infty.$ In this case we have that $b\rightarrow0$,
which means that the $n$ points are uniformly and independently distributed
on the straight line. Let us now consider a disk of radius $r>0$
centered at each of these points and let us connect every point to
the other points which lie inside its disk. Thus, the resulting graph
resembles a one-dimensional RGG, that is a graph created by placing
the $n$ points uniformly and independently on the interval $\left[0,1\right]$
and then connecting pairs of nodes if they are at a (Euclidean) distance
smaller than or equal to a certain connection radius $r$ (see for
instance \cite{One-dimensional RGG_1,One-dimensional RGG_2,One-dimensional RGG_3}).

\section{Topological Properties of RRGs}

\subsection{Average degree}

We start the study of the topological properties of RRGs by considering
an analytical expression for the average degree $\bar{k}$. We remind
that the degree of a node is the number of edges connected to it.
The average degree is a property not only interesting by itself but
as we will see in the next sections of this paper it is fundamental
to understand other topological parameters of RRGs.

\noindent To start with, let us consider that for a given node, there
are $n-1$ nodes distributed in the rest of the rectangle. Define
$A_{p}$ to be the area within the radius $r$ of a point $p$ which
lies within the rectangle. Since the nodes are uniformly and independently
distributed, the expected node degree of a node $v_{i}$ is $\mathbf{E}(k_{i})=(n-1)A_{i}/(ab)$,
where $A_{i}$ is taken for the point where node $v_{i}$ is located.
This is because dividing the nodes between the area within distance
$r$ and the rest of the rectangle gives rise to the Binomial distribution
$Bin(n-1,A_{i}/(ab))$. Averaging this over all possible node locations
(i.e., the points in the rectangle) gives

\noindent 
\begin{equation}
\mathbf{E}\bar{k}=\dfrac{\int_{p}\lbrace(n-1)A_{p}/(ab)\rbrace}{ab}=\dfrac{(n-1)\int_{p}A_{p}}{(ab)^{2}}.
\end{equation}

Let $f(a,b,r)$ be the area within radius $r$ of a point which lies
in the rectangle, integrated over all points, i.e., $f(a,b,r)=\int_{p}A_{p}$.
Based on preliminary computational results (not shown) obtained for
the average degree we consider here the following three regions: $0\leq r\leq b$,
$b\leq r\leq a$ and $a\leq r\leq\sqrt{a^{2}+b^{2}}$, recalling that
$a\geq b$. We call these cases 1, 2 and 3, respectively. Thus, the
function $f(a,b,r)$ takes different forms $f_{i}$ for each case
$i$. This means that we can write 

\begin{equation}
\mathbf{E}\bar{k}=\dfrac{(n-1)f_{i}}{(ab)^{2}},\label{eq:expected degree}
\end{equation}

with

\begin{equation}
f_{i}=\begin{cases}
f_{1} & 0\leq r\leq b,\\
f_{2} & b\leq r\leq a,\\
f_{3} & a\leq r\leq\sqrt{a^{2}+b^{2}},
\end{cases}
\end{equation}

and our task is now to find the analytical expressions for $f_{i}$.

\begin{figure}[t]
\centering \includegraphics[width=0.75\textwidth]{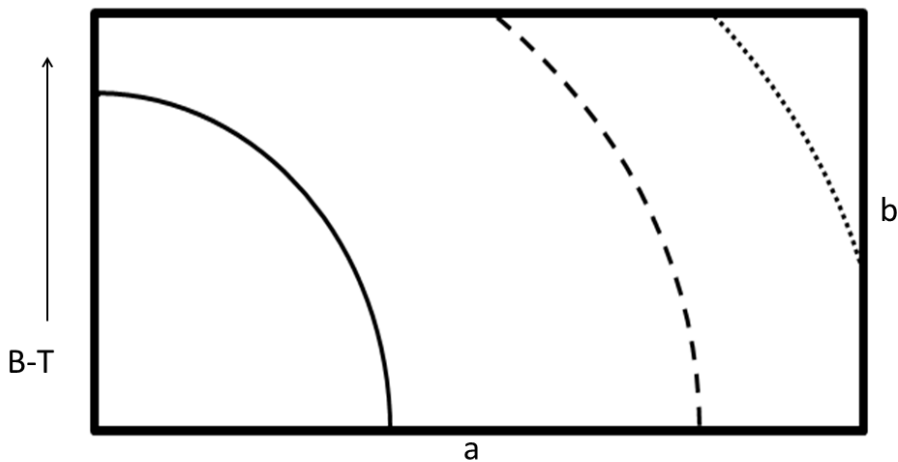}

\protect\caption{Illustration of three different quarter circles in the rectangle corresponding
to $0\leq r\leq b$ (solid line), $b\leq r\leq a$ (broken line) and
$a\leq r\leq\sqrt{a^{2}+b^{2}}$ (dotted line). The direction used
for displacing the circles is represented as bottom-top (B-T) with
an arrow in the graphic. }

\label{rectangle} 
\end{figure}

We consider the rectangle in Fig. \ref{rectangle}, which shows 3
quarter circles of different radii (each corresponding to one of the
three cases) as they intersect the interior of the rectangle. We consider
only quarter circles instead of circles, then quadruple the result
at the end. For each of these quarter circles, we divide them into
vertical rectangular strips of width $\Delta x$ which will approximate
the areas of the intersection between the quarter circles and the
full rectangle; this will become exact in the limit as $\Delta x\rightarrow0$.
We now consider several possibilities for these strips.

First, the strips may approximate an area which is not rectangular,
which occurs when the height of the strip is smaller than the height
$b$ of the rectangle. For a strip of distance $x$ from the left
of the rectangle, this corresponds to $0\leq x\leq r$ for the smaller
quarter circle (case 1), $\sqrt{r^{2}-b^{2}}\leq x\leq r$ for the
medium quarter circle (case 2), and $\sqrt{r^{2}-b^{2}}\leq x\leq a$
for the largest quarter circle (case 3). Setting $p=min(a,r)$, $q=min(b,r)$,
we have $\sqrt{r^{2}-q^{2}}\leq x\leq p$.

Since we need to integrate these areas over all possible quarter circles,
given a fixed radius, we wish to know how we can translate the quarter
circle in the figure and preserve a particular strip. That is, for
a particular strip of distance $x$ from the left of the rectangle,
we may find a corresponding strip on the other quarter circles of
the same radius. Since we have a rectangular strip of width $\Delta x$,
height $\sqrt{r^{2}-x^{2}}$, and distance $x$ from the center of
the (quarter) circle, we may find this strip in any of the $(a-x-\Delta x)$
positions horizontally, and $(b-\sqrt{r^{2}-x^{2}})$ vertically.
Thus, we can use integration to find the total area of all these strips
by multiplying $(a-x-\Delta x)(b-\sqrt{r^{2}-x^{2}})$ by the area
of the strip $\sqrt{r^{2}-x^{2}}\,\Delta x$, and taking the limit
to obtain

\begin{align}
I_{1} & =\int_{\sqrt{r^{2}-q^{2}}}^{p}(a-x)(b-\sqrt{r^{2}-x^{2}})\sqrt{r^{2}-x^{2}}\, dx\nonumber \\
 & =\int_{\sqrt{r^{2}-q^{2}}}^{p}(a-x)(b\sqrt{r^{2}-x^{2}}-(r^{2}-x^{2}))\, dx.
\end{align}

Secondly, we note that if these strips are translated far enough in
the bottom-top (B-T) direction, they become truncated by the top of
the rectangle. For a particular truncated strip, we may still find
a corresponding strip in any of the ($a-x-\Delta x$) positions horizontally,
and the truncated height $t$ of a strip may be any value between
0 and the full height of the strip. Thus, integrating gives

\begin{align}
I_{2} & =\int_{\sqrt{r^{2}-q^{2}}}^{p}(a-x)\int_{0}^{\sqrt{r^{2}-x^{2}}}t\, dt\, dx\nonumber \\
 & =\int_{\sqrt{r^{2}-q^{2}}}^{p}\dfrac{1}{2}(a-x)(r^{2}-x^{2})\, dx.
\end{align}

Alternatively, we may have $\sqrt{r^{2}-b^{2}}>b$, in which case
the rectangular strip is exact and of height $b$. In this case, the
only contribution is from the truncated strips. We note that this
applies for $0\leq x\leq\sqrt{r^{2}-q^{2}}$ by a similar argument
as before, and we integrate as follows

\begin{align}
I_{3} & =\int_{0}^{\sqrt{r^{2}-q^{2}}}(a-x)\int_{0}^{b}t\, dt\, dx\nonumber \\
 & =\int_{0}^{\sqrt{r^{2}-q^{2}}}\dfrac{1}{2}(a-x)b^{2}\, dx.
\end{align}

Thus, we have the expression for $f$ as four times the sum of the
above integrals

\begin{align}
f & =4(I_{1}+I_{2}+I_{3})\nonumber \\
 & =\int_{0}^{\sqrt{r^{2}-q^{2}}}2(a-x)b^{2}dx+\int_{\sqrt{r^{2}-q^{2}}}^{p}(a-x)(4b\sqrt{r^{2}-x^{2}}-2(r^{2}-x^{2}))dx,
\end{align}

which can be written as

\begin{equation}
f=\begin{cases}
0\leq r\leq b & \pi r^{2}ab-\frac{4}{3}(a+b)r^{3}+\frac{1}{2}r^{4},\\
b\leq r\leq a & -\frac{4}{3}ar^{3}-r^{2}b^{2}+\frac{1}{6}b^{4}+a(\frac{4}{3}r^{2}+\frac{2}{3}b^{2})\sqrt{r^{2}-b^{2}}\\
 & +2r^{2}ab\arcsin(\frac{b}{r}),\\
a\leq r\leq\sqrt{a^{2}+b^{2}} & -r^{2}(a^{2}+b^{2})+\frac{1}{6}(a^{4}+b^{4})-\frac{1}{2}r^{4}\\
 & \quad+b(\frac{4}{3}r^{2}+\frac{2}{3}a^{2})\sqrt{r^{2}-a^{2}}+a(\frac{4}{3}r^{2}+\frac{2}{3}b^{2})\sqrt{r^{2}-b^{2}}\\
 & \quad-2abr^{2}(\arccos(\frac{b}{r})-\arcsin(\frac{a}{r})).
\end{cases}\label{eq:expected values}
\end{equation}

\begin{figure}[h!]
\subfigure[]{}\includegraphics[width=0.5\textwidth]{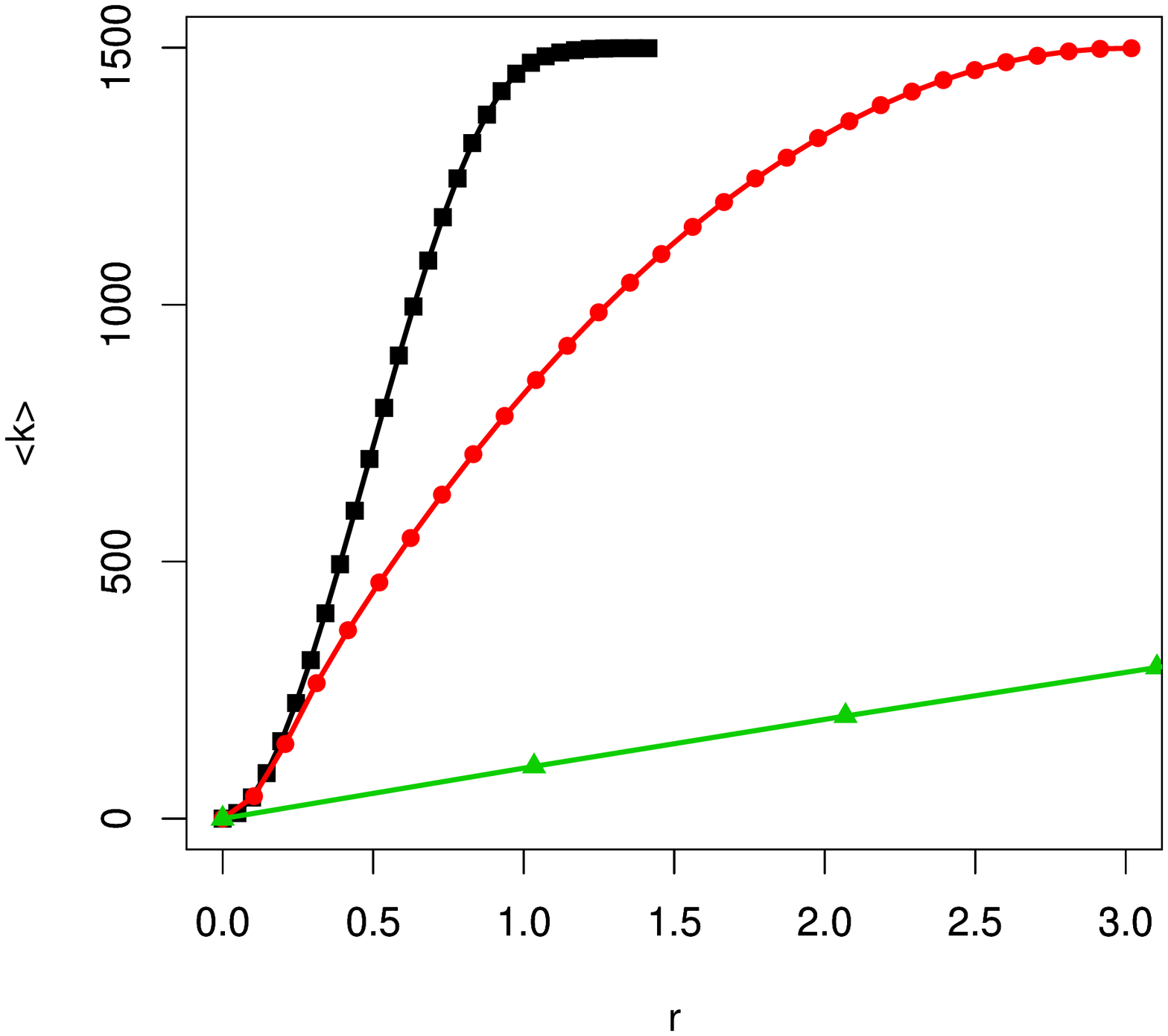}\subfigure[]{}\includegraphics[width=0.5\textwidth]{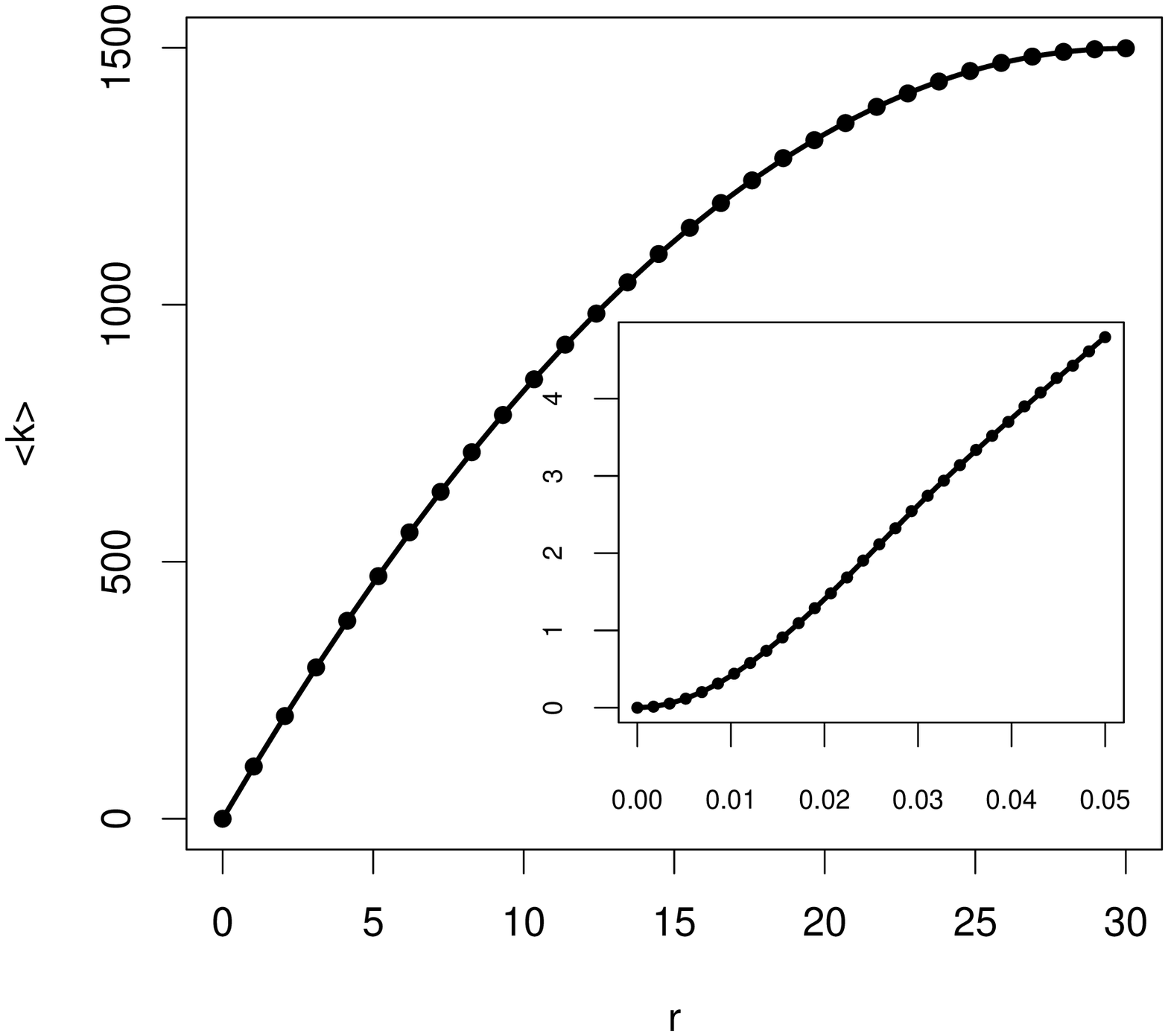}

\protect\protect\protect\caption{(color online) (a) Illustration of the fit between the observed (black
squares ($a=1$), red circles ($a=3$), and green triangles ($a=30$))
and expected (solid line) values of the average degree for RRGs with
different side lengths of the rectangle. (b) Wider range of radii
for $a=30$ (zooming for the small radii in the inset). }

\label{Observed vs. expected} 
\end{figure}

Now we evaluate computationally how good the expression for the average
degree of the RRGs is. In Fig. \ref{Observed vs. expected}(a) we
plot the values of the average degree observed for RRGs with three
different values of the rectangle side length. These observed values
(represented by solid squares, circles and triangles) are the average
of 100 random realizations of RRGs with $n=1,500$ nodes. The solid
lines represent the expected values according to the expressions (\ref{eq:expected values}).
The Pearson correlation coefficients for the linear regression between
the observed and expected values are larger than 0.9999 in the three
cases. We enlarge the region of small radii for the case $a=30$ (see
Fig. \ref{Observed vs. expected} (b)) where it can be seen that it
is a perfect fit also for this region with Pearson correlation coefficient
as good as for the general case.

\subsection{Connectivity}

In this section we are interested in determining the connection radius
for specific values of $n$, i.e., $r\left(n\right)$, which guarantees
that the RRG $\Gamma\left(n,r\left(n\right)\right)$ is asymptotically
connected with probability one. For the RGG, Penrose \cite{Penrose connectivity}
proved that if $M_{n}$ is the maximum length of an edge in the graph,
then the probability that $n\pi M_{n}^{2}-\log n\leq\alpha$ for a
given $\alpha\in\mathbb{R}$ is

\begin{equation}
\lim_{n\rightarrow\infty}P\left[n\pi M_{n}^{2}-\log n\leq\alpha\right]=\exp\left(-\exp\left(-\alpha\right)\right).
\end{equation}

In 2D $M_{n}=r$ , such that we have for the RGG

\begin{equation}
\lim_{n\rightarrow\infty}P\left[n\pi r^{2}-\log n\leq\alpha\right]=\exp\left(-\exp\left(-\alpha\right)\right).
\end{equation}

This means that for $\alpha\rightarrow+\infty$ the RGG is almost
surely connected when $n\rightarrow\infty$, and almost surely disconnected
when $\alpha\rightarrow-\infty$. 

The term $n\pi r^{2}$ is just the average degree $\bar{k}$ in RGG
(when no boundary effects are considered). Thus we can write

\begin{equation}
\lim_{n\rightarrow\infty}P\left[\bar{k}-\log n\leq\alpha\right]=\exp\left(-\exp\left(-\alpha\right)\right).\label{eq:ProbConnectivity}
\end{equation}

In the previous section we have obtained an analytic expression for
$\bar{k}$ in the RRG. Thus, we can replace this in (\ref{eq:ProbConnectivity})
and obtain an analogous expression for the RRG. That is, for a RRG
we have that

\begin{equation}
\lim_{n\rightarrow\infty}P\left[\dfrac{(n-1)f_{i}}{(ab)^{2}}-\log n\leq\alpha\right]=\exp\left(-\exp\left(-\alpha\right)\right),\label{eq:ConnecRRG}
\end{equation}

where $f_{i}$ is given by (\ref{eq:expected values}).

Because the parameter $\alpha$ is unknown and it depends on the specific
RRG considered, we can obtain a lower bound for $\exp\left(-\exp\left(-\alpha\right)\right)$
using (\ref{eq:ConnecRRG}). That is,

\begin{equation}
\exp\left(-\exp\left(-\left(\dfrac{(n-1)f_{i}}{(ab)^{2}}-\log n\right)\right)\right)\leq\exp\left(-\exp\left(-\alpha\right)\right).\label{eq:lowerbound connectivity}
\end{equation}

\begin{figure}
\subfigure[]{}\includegraphics[width=0.33\textwidth]{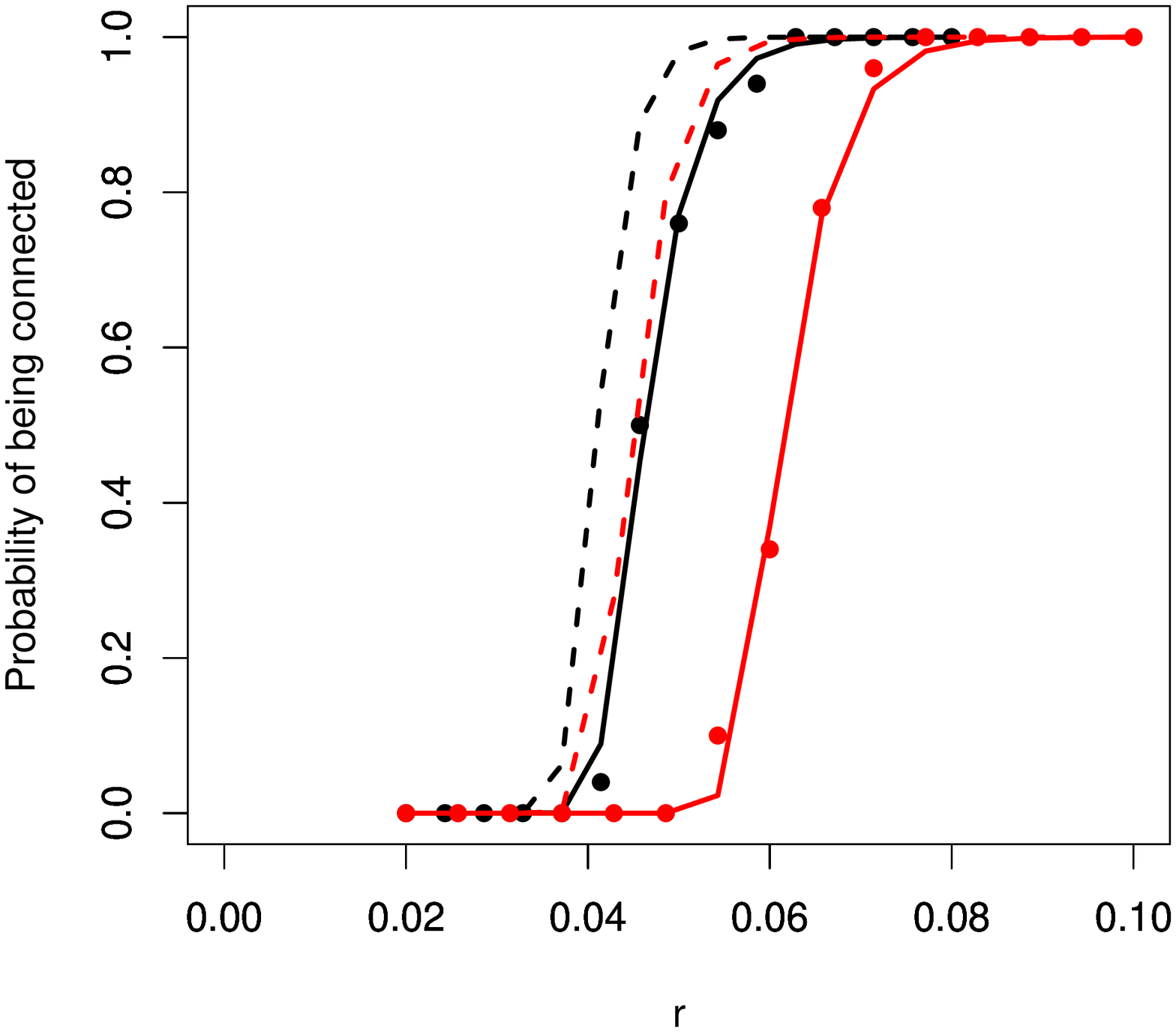}\subfigure[]{}\includegraphics[width=0.33\textwidth]{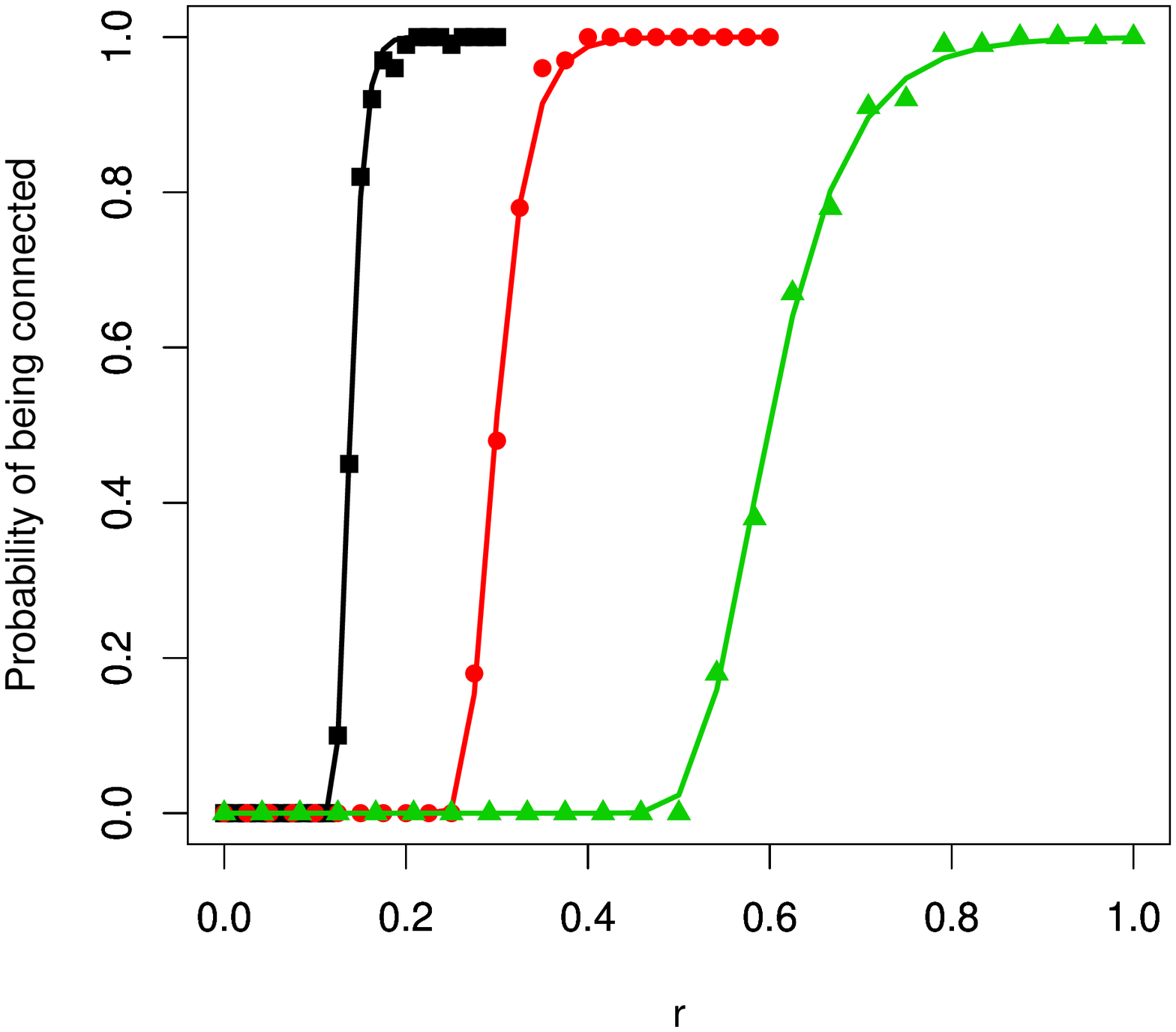}\subfigure[]{}\includegraphics[width=0.33\textwidth]{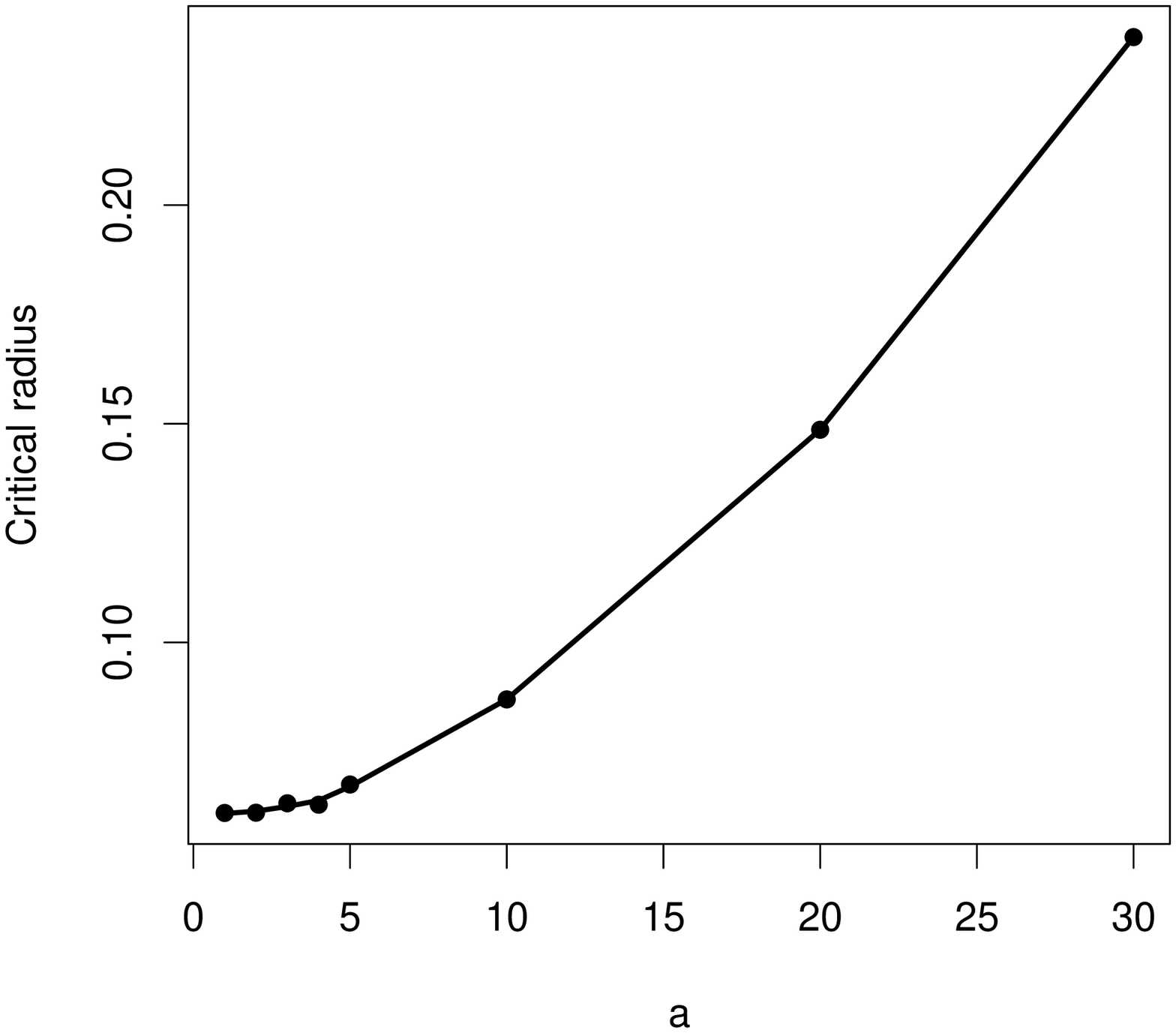}

\protect\caption{Dependence of the probability of being connected with the connection
radius in RRGs. a) Illustration of the goodness of the bound (\ref{eq:lowerbound connectivity})
for two RRGs with $a=1$ (black curves), and $a=5$ (red curves).
The observed values for RRGs with 1,500 nodes averaged over 100 random
realizations are illustrated as solid circles and the values obtained
by the theoretical bound are given by broken lines. (b) Observed values
of the probability of being connected for RRGs with 1,500 nodes averaged
over 100 random realizations and three different rectangle side lengths
$a=1$ (black squares), $a=5$ (red circles), and $a=10$ (green triangles).
(c) Variation of the radius for which the RRG is connected (critical
radius) with the rectangle side length for RRGs with 1,500 nodes and
11,250 edges.}

\label{connectivity}
\end{figure}

Now we consider the computational evaluation of the connectivity of
RRGs with 1,500 nodes as a function of the connection radius and the
rectangle side length. We start by analyzing the goodness of the upper
bound that we have found for the probability of a RRG to be connected
(see (\ref{eq:lowerbound connectivity})). In Fig. \ref{connectivity}(a)
we illustrate the plot of the probability of being connected versus
the connection radius for graphs embedded into rectangles of side
length $a=1$ (black circles) and $a=10$ (red circles). The solid
circles represent the average values of 100 random realizations for
these graphs. The values corresponding to the upper bound are plotted
as broken lines. As can be seen, both the observed and the upper bound
follow the same distribution and the upper bound is relatively close
to the average observed values. In Fig. \ref{connectivity}(b) we
illustrate the change in the probability that the RRG is connected
as a function of the connection radius, i.e., the minimum radius needed
to make the network connected, for three different values of $a.$
As the square is elongated the critical radius increases with the
value of $a$. For instance, for $a=1$ the critical radius is about
0.2, and for $a=10$ it is about 0.9. Then, for $a=1$, $\alpha\geq185.3$
for the RRG to be connected. This value increases up to $\alpha\geq750.8$
for $a=5$ and to $\alpha\geq3813.8$ for $a=10$. The main reason
for this increase in the critical radius is that as we elongate the
rectangle the points have to cover a longer region and consequently
their separation increases. As a consequence, we need to increase
the connection radius in order to guarantee the connectivity of the
network. In other words, increasing the value of $a$ necessarily
implies having to increase the connection radius to make the network
connected. The global increase of the critical radius with the side
length of the rectangle is given in Fig. \ref{connectivity} (c)\textcolor{black}{. }

\subsection{Degree distribution}

In the RRG the $n$ nodes are distributed uniformly and independently
on the unit rectangle. Then, the degree distribution can be easily
estimated by considering the probability density function of having
a node $i$ of degree $k_{i}$ given that there are $n-1$ other nodes
uniformly distributed in the unit rectangle (see for instance \cite{Degree distribution}).
This gives rise to the binomial distribution of the node degrees,
which when $\left(n-1\right)\sim n$ is given by

\begin{equation}
p\left(k\right)=\left(\begin{array}{c}
n\\
k
\end{array}\right)p^{k}\left(1-p\right)^{n-k}.
\end{equation}

When $n\rightarrow\infty$ and $p$ is sufficiently small, this binomial
distribution approaches very well a Poisson distribution of the form

\begin{equation}
p\left(k\right)\simeq\dfrac{\bar{k}^{k}\exp\left(-\bar{k}\right)}{k!}.\label{eq:Poisson distribution}
\end{equation}

\begin{figure}
\begin{centering}
\includegraphics[width=0.75\textwidth]{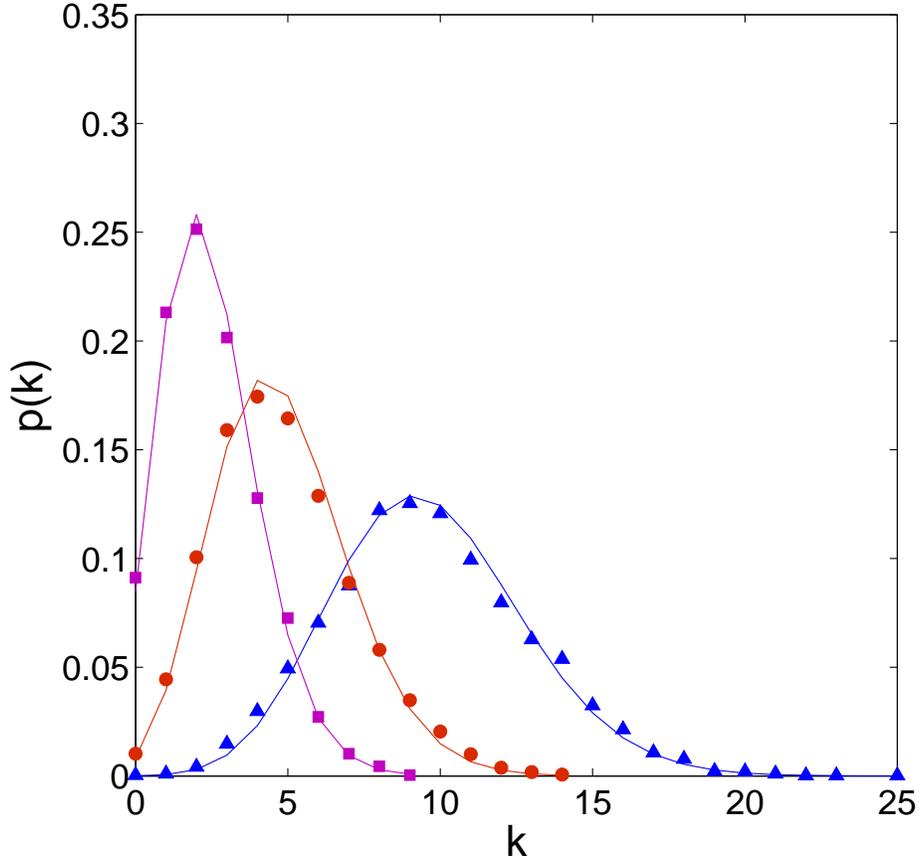}
\par\end{centering}

\protect\caption{Degree distribution of RRGs with $n=5,000$, connection radius $r=0.025$,
and rectangle side lengths $a=1$ (triangles), $a=50$ (circles),
$a=100$ (squares). The solid figures (triangles, circles and squares)
correspond to the average of 100 random realizations for the given
network. The solid lines correspond to the shape of the Poisson distribution
(\ref{eq:Poisson distribution}) with the corresponding average degree
obtained from eq. (\ref{eq:expected values}). }

\label{degree distributions}
\end{figure}

As we have previously obtained an analytic expression for the average
degree $\bar{k}$ we can easily compute the degree distribution for
RRGs. We select RRGs with 5,000 nodes and radius of connection equal
to 0.025. Then, we obtain the degree distribution for different values
of the rectangle side length and take the average of 100 random realizations.
In Fig. \ref{degree distributions} we also plot the expected distribution
using the equation (\ref{eq:Poisson distribution}) in which we have
plugged the values of the expected average degree obtained previously.
As can be seen, independently of the side length of the rectangle,
the RRG displays Poisson degree distributions. That is, the elongation
of the rectangle does not affect the shape of the degree distribution
of the nodes. 

\subsection{Average shortest path distance}

Let $\Gamma=\left(V,E\right)$ be a simple connected graph. A \textit{path}
of length $k$ in $\Gamma$ is a set of nodes $i_{1},i_{2},\ldots,i_{k},i_{k+1}$
such that for all $1\leq l\leq k$, $(i_{l},i_{l+1})\in E$ with no
repeated nodes. The \textit{shortest-path }or \textit{geodesic distance}
between two nodes $u,v\in V$ is defined as the length of the shortest
path connecting these nodes. We will write $d(u,v)$ to denote the
distance between $u$ and $v$. Here we will define, as is usual in
network theory, the average path length to be the following quantity:

\begin{equation}
\left\langle l\right\rangle =\dfrac{2}{n\left(n-1\right)}\sum_{u<v}d\left(u,v\right).
\end{equation}

The diameter of the graph is defined as the maximum of all the shortest
path lengths in the graph, i.e., $D=\max_{u,v}d\left(u,v\right).$
We consider here an upper bound for the average path length. Then,
we start by considering that the $n$ points are distributed homogeneously
in the rectangle of sides $a$ and $b=a^{-1}$ in such a way that
the points are equally spaced in the rectangle and are separated by
a Euclidean distance $r$. In this case, the point which is the farthest
from the rest of the $n-1$ points in the rectangle is one at any
of the four corners of the rectangle. Let us designate this point
as $i$. If the average path length of this node is $\left\langle l_{i}\right\rangle $,
it is straightforward to realize that $\left\langle l\right\rangle \leq\left\langle l_{i}\right\rangle $,
and we will have the desired upper bound. It is easy to see that the
largest chain of connected points is realized along the main diagonal
of the rectangle. Thus, the longest distances involving the node $i$
are those connecting it with the other nodes $j$ along the main diagonal
of the rectangle. Let us designate the length of the main diagonal
is $c$. Then, there are $D=\dfrac{c}{r}$ connected nodes in this
line. Let us now consider $\left\langle l_{i}\right\rangle $ based
only on those $j$ nodes along the main diagonal of the rectangle
(notice that this is an upper bound for $\left\langle l_{i}\right\rangle $).
That is

\[
\sum_{j\in diag}d_{ij}=1+2+\cdots+D=\dfrac{D\left(D+1\right)}{2}.
\]

Consequently,

\[
\left\langle l_{i}\right\rangle \leq\dfrac{D\left(D+1\right)}{2D}=\dfrac{D+1}{2}.
\]

Because, $\left\langle l\right\rangle \leq\left\langle l_{i}\right\rangle $,
we have

\[
\left\langle l\right\rangle \leq\left\langle l_{i}\right\rangle \leq\dfrac{D+1}{2}.
\]

It is easy to note that $D<\dfrac{a}{r}+\dfrac{1}{ar}$. That is,
the length of the diagonal is smaller than the sum of the length of
the two sides of the rectangle (in terms of number of nodes), which
are $\dfrac{a}{r}$ and $\dfrac{b}{r}=\dfrac{1}{ar}$. Then, we have

\begin{equation}
\left\langle l\right\rangle \leq\left\langle l_{i}\right\rangle \leq\dfrac{D+1}{2}\leq\dfrac{\dfrac{a}{r}+\dfrac{1}{ar}+1}{2}=\dfrac{a^{2}+ar+1}{2ar}.\label{eq:distance bound}
\end{equation}

An important consequence of this bound is that it allows us to find
the asymptotic \foreignlanguage{american}{behavior} of the average
path length as the rectangle becomes more elongated. That is, if we
fix the connection radius we can see that $\lim_{a\rightarrow\infty}\left\langle l\right\rangle =\infty$.
In other words, as the rectangle becomes more elongated the RRG becomes
a large-world with a very large average shortest path distance. As
$a\rightarrow\infty$, the connection radius $r$ has to be increased
to keep the connectivity of the RRG. Thus, the infinite growth of
the path length is not observed in practice due to the connectivity
constraints imposed by the connection radius. Our further calculations
show that the average path length is of the same order of magnitude
as the right-hand side of eq. (\ref{eq:distance bound}), which means
that we can write it as:

\begin{equation}
\left\langle l\right\rangle \sim\dfrac{a^{2}+ar+1}{2ar}.\label{eq:distance bound-1}
\end{equation}

\begin{figure}
\subfigure[]{}\includegraphics[width=0.33\textwidth]{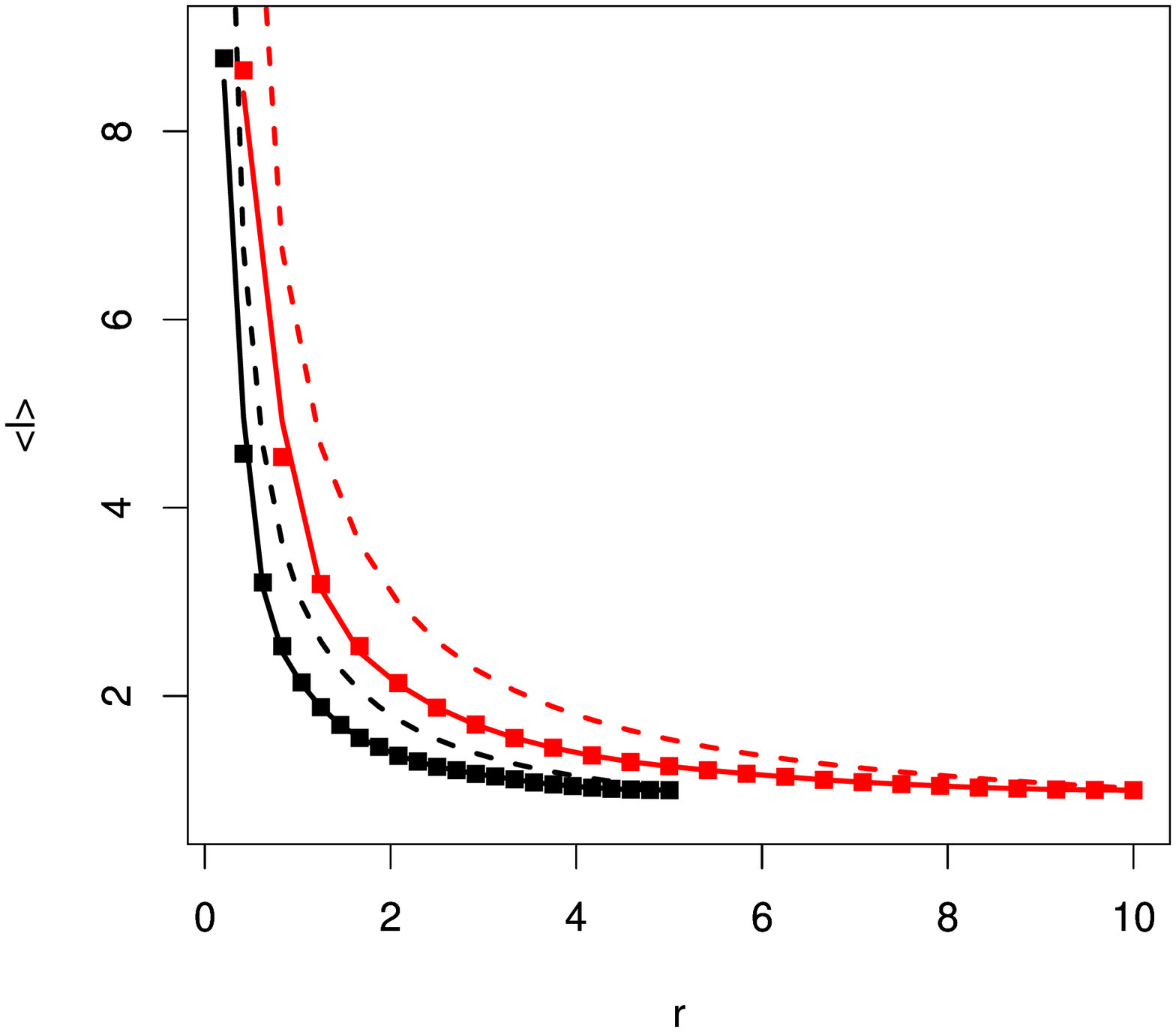}\subfigure[]{}\includegraphics[width=0.33\textwidth]{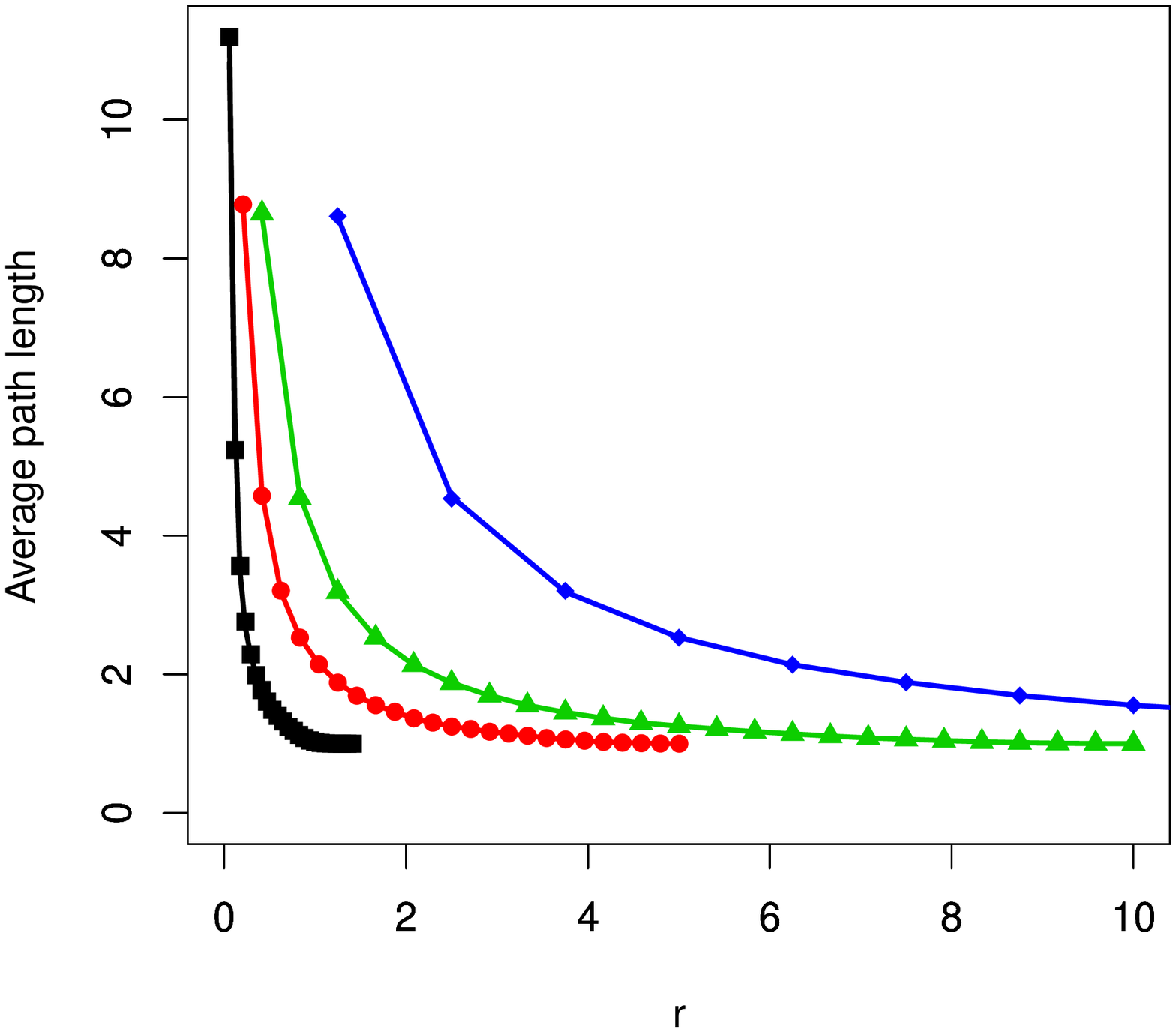}\subfigure[]{}\includegraphics[width=0.33\textwidth]{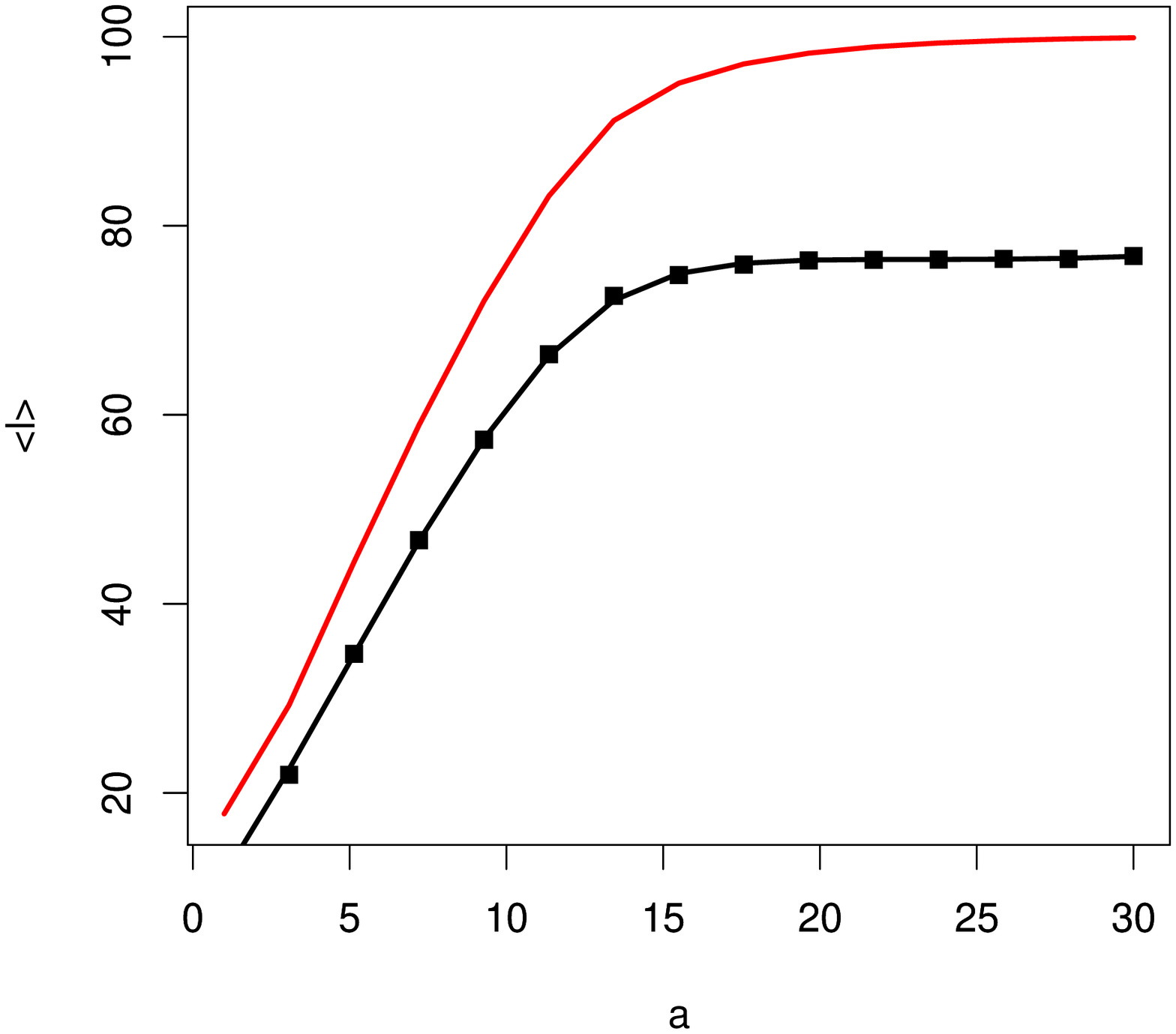}

\protect\caption{(a) Plot of the average path length for RRGs with $n=1,500$ nodes
with different connection radii for $a=5$ (black squares) and for
$a=10$ (red squares). The solid squares represent the average of
$\left\langle l\right\rangle $ for 100 random realizations (lines
connecting the squares are used to guide the eye). The dotted lines
represent the upper bounds (\ref{eq:distance bound}) for $\left\langle l\right\rangle $
in RRGs for the corresponding values of $a$. (b) Change of the average
path length versus radius for networks with $n=1,500$ nodes for different
values of $a$: black squares ($a=1$), red circles ($a=5$), green
triangle ($a=10$), blue rhombus ($a=30$). (c) Illustration of the
upper bound (solid line) for the average path length as a function
of $a$ for a network with the same size as before and connection
radii that guarantees $m=11,250$ edges for a given value of $a$.}

\label{average path length}
\end{figure}

We now study computationally RRGs with 1,500 nodes. For every value
of $a$ we report the average of 100 random realizations. First, we
analyze the goodness of the upper bound found here for the average
path length. In Fig. \ref{average path length}(a) we illustrate the
variation of $\left\langle l\right\rangle $ with the connection radius
for two different values of $a,$ namely $a=5$ and $a=10$. In the
same plot we illustrate the values of the upper bound obtained with
the expression (\ref{eq:distance bound}), where it can be seen that
the upper bound is very close to the average shortest path obtained
for these RRGs. Particularly, for large values of $r$ the observed
values are almost identical to those of the upper bound. As observed
in Fig. \ref{average path length}(b) the average path length not
only changes with the variation of the connection radius but also
with the rectangle side length. This is already expected from the
eq. (\ref{eq:distance bound-1}) where it can be seen that as $a\rightarrow\infty$
the average path length also grows to infinity for fixed $r$ (see
observation at the end of the previous paragraph). 

These results agree with our intuition that as we elongate the rectangle
there are nodes which are farther apart from each other and as a result
the average path length of the whole graph increases. For a better
analysis of this relation we plotted the results of the average path
length for 100 random realizations of the previously mentioned RRGs
versus $a$, with $m=11,250$ edges, in Fig. \ref{average path length}(c).
We also plot here the upper bound (\ref{eq:distance bound}). It can
be seen that there is an almost linear increase of $\left\langle l\right\rangle $
for values of $1\leq a\lesssim15$ after which the dependence is very
flat. In this region we have that $a\rightarrow\infty$, which corresponds
to a good approximation of a one-dimensional RGG. For $a=1$ it is
known that the average path length depends only on the inverse of
the radius, $\left\langle l_{e}\right\rangle =\varTheta\left(1/r\right)$.
The actual radius used for the plot in Fig. \ref{clustering vs a}
(c) is\textcolor{black}{{} $r=0.0578$, which gives an estimate of the
average path length of 17.3, which is not too far from} the observed
value in the plot for $a=1$. In the case of $a=30$ we are in the
presence of a very elongated rectangle, which is very similar to a
one-dimensional RGG. A crude estimate of the average path length in
this case would be $\left\langle l_{e}\right\rangle =n/\left\langle k\right\rangle $,
which in the current case will give $\left\langle l_{e}\right\rangle \approx100$,
which is relatively close to the observed value of $\left\langle l\right\rangle \approx75$
for $a=30$.

\subsection{Clustering coefficient}

One of the most important network parameters is the \textit{local
clustering coefficient, }usually known as the \textit{Watts-Strogatz
clustering coefficient,} of a node $i$. This parameter is defined
as \cite{Watts-Strogatz}:
\begin{equation}
C_{i}=\frac{2t_{i}}{k_{i}(k_{i}-1)},
\end{equation}

where $t_{i}$ is the number of triangles involving the node $i$
and $k_{i}$ is the degree of the node $i$. Taking the mean of these
values as $i$ varies among the nodes in $\Gamma$, one gets the \textit{average
clustering coefficient} of the network: $\left\langle C\right\rangle =\frac{1}{n}\sum_{i=1}^{n}C_{i}.$

The average clustering coefficient of a RGG has been obtained by Dall
and Christensen \cite{Dall Christense} for $r^{2}=\dfrac{\log n+\alpha}{n\pi}$
when $n\rightarrow\infty$ and $\alpha\rightarrow\infty$, where $\alpha\in\mathbb{R}$
is a constant for a given number of nodes (see Section B. Connectivity):

\begin{equation}
\left\langle C_{d}\right\rangle =\left\{ \begin{array}{cc}
1-H_{d}\left(1\right) & d\ even\\
\frac{3}{2}H_{d}\left(1/2\right) & d\ odd,
\end{array}\right.
\end{equation}

where $d$ is the dimension of the hypercube in which the nodes are
embedded and

\begin{equation}
H_{d}\left(x\right)=\dfrac{1}{\sqrt{\pi}}\sum_{i=x}^{d/2}\dfrac{\Gamma\left(i\right)}{\Gamma\left(i+\frac{1}{2}\right)}\left(\dfrac{3}{4}\right)^{i+\frac{1}{2}},
\end{equation}

where $\Gamma\left(i\right)$ is the Gamma function. Thus, for $d=2$
, $\left\langle C_{2}\right\rangle =1-\cfrac{3\sqrt{3}}{4\pi}\approx0.5865$
and for $d=1$, $\left\langle C_{1}\right\rangle =3/4=0.75$.

Here, however, we are interested in an expression that accounts for
the variations of the clustering coefficient with both the connection
radius and the rectangle side length. Our strategy is similar to the
one used in \cite{Dall Christense}. That is, let $i$ and $j$ be
two connected nodes in a RRG, which are separated at a Euclidean distance
$\delta$ from each other. Let us draw two circles of radius $r$
centered respectively at $i$ and $j$. Let $\delta\leq r$ such that
the two nodes are connected. Then, because $\delta<2r$ the two circles
overlap. Because $i$ and $j$ are connected, any point in the area
formed by the overlap of the two circles will form a triangle with
the nodes $i$ and $j$ . In addition, any node inside the two circles
which is not in the overlapping area forms a path of length two with
the nodes $i$ and $j$ . Thus if we quantify the ratio of the overlapping
area to the total area of the circle we account for the ratio of the
number of triangles to open triads in which the nodes $i$ and $j$
take place, i.e., the clustering coefficient. This ratio is given
by

\begin{equation}
\left\langle C\right\rangle =\dfrac{2r^{2}\arccos\left(\dfrac{\delta}{2r}\right)-\dfrac{1}{2}\delta\sqrt{4r^{2}-\delta^{2}}}{\pi r^{2}}.\label{eq:clustering}
\end{equation}

At this point we only need an estimation of the length $\delta$ between
two connected nodes in a RRG. We use here a simple approach based
on the following intuition. Let us start by considering $n$ nodes
in a square in such a way that they form a regular square lattice.
Then, $\delta$ is proportional to the length of the side of the rectangle
$a$ divided by the number of circles along this side. As we have
a square, the number of points along the side of length $a$ is the
same as that for the other side. Consequently, $\delta\sim n^{-1/2}$.
If we elongate the rectangle to $a\rightarrow\infty$, which resembles
a straight line, we will have that the separation between the two
points is just the length of the straight line divided by the number
of nodes, $\delta\sim an^{-1}$. For a general rectangle the separation
between two points in a line along the edge side of length $a$ is
given by $\delta$$\sim an^{-\gamma}$, where $\gamma\sim a\left(a+b\right)^{-1}$.
Notice that when $a=1$ ($b=1$) we have $\delta\sim n^{-1/2}$ and
when $a\rightarrow\infty$ ($b\rightarrow0$) we have $\delta\sim an^{-1}$.
We remind the reader that here we consider only the case $b=a^{-1}$.

\begin{figure}[h!]
\subfigure[]{}\includegraphics[width=0.33\textwidth]{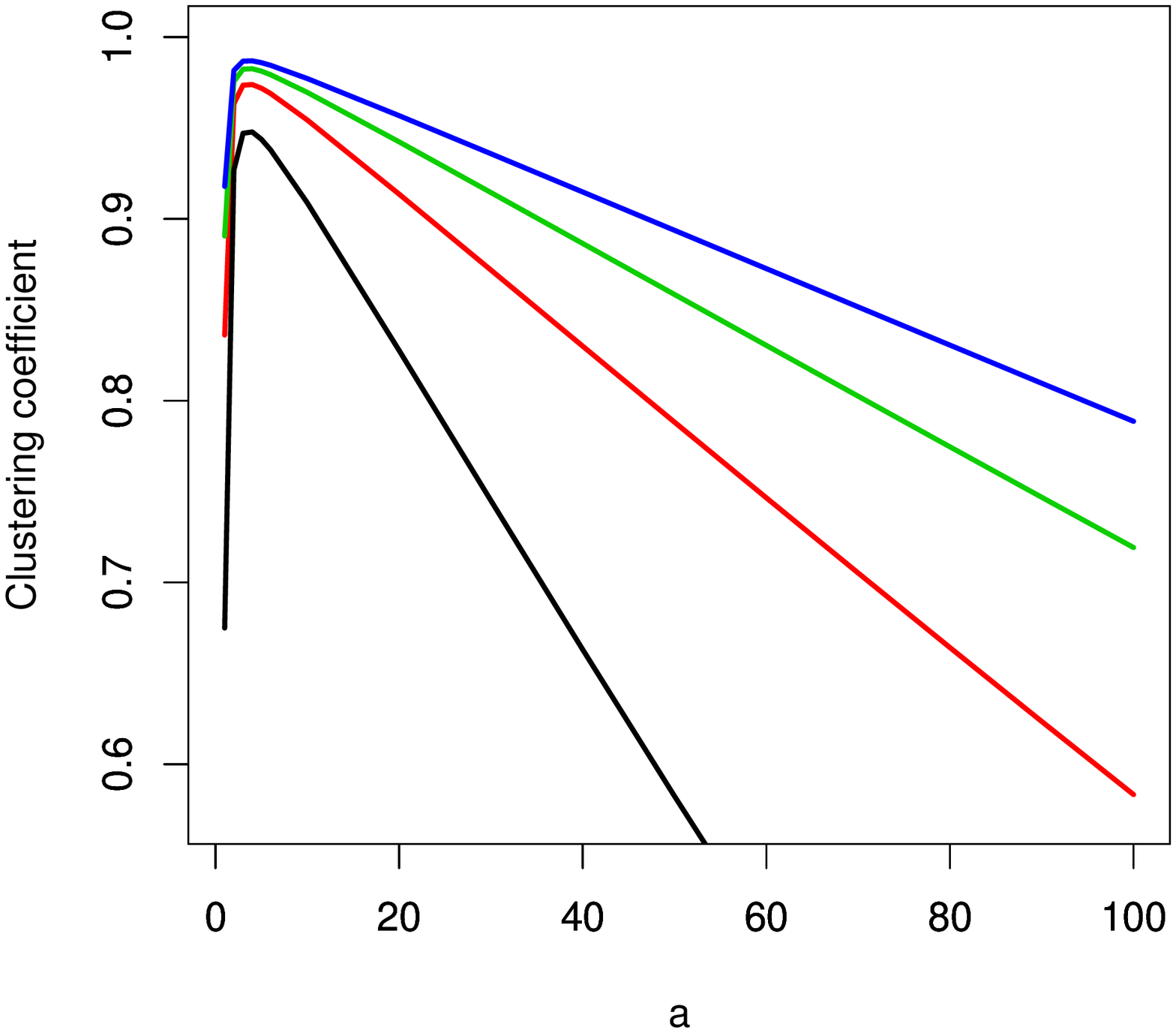}\subfigure[]{}\includegraphics[width=0.33\textwidth]{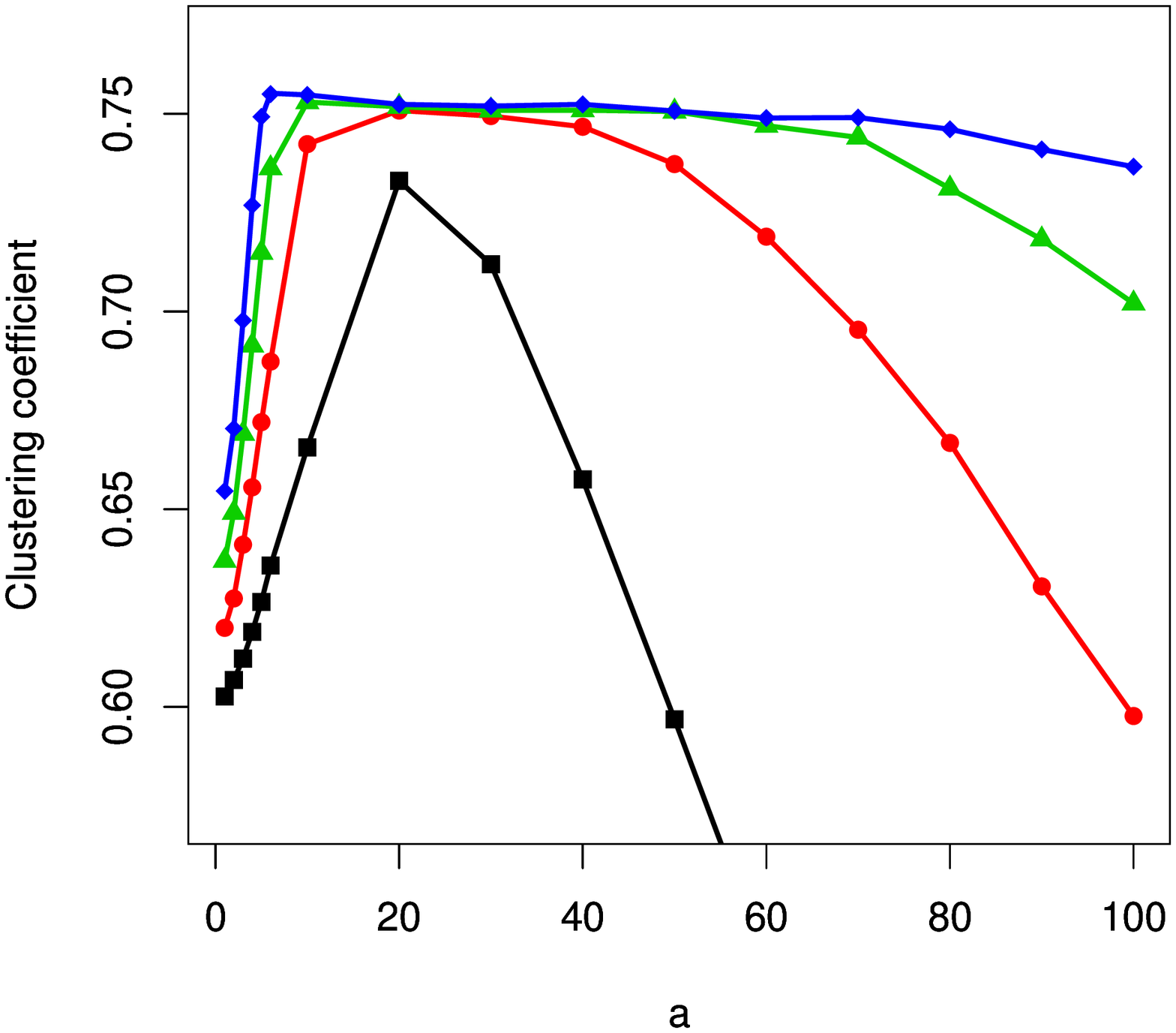}

\protect\caption{Illustration of the dependence of the clustering coefficient with
the rectangle side length for different connection radii: $r=0.05$
(black), $r=0.1$ (red), $r=0.15$ (green) $r=0.20$ (blue). In plot
(a) we show the analytical results and in plot (b) the observed ones.
Both results are obtained for RRG with 1,500 nodes and for the observed
ones we averaged the results of 100 random realizations. }

\label{clustering vs a} 
\end{figure}

Plugging $\delta\sim an^{-\gamma}$, with $\gamma\sim a\left(a+b\right)^{-1}$
into (\ref{eq:clustering}) we obtain some surprising results. In
Fig. \ref{clustering vs a}(a) we illustrate the dependence of $\left\langle C\right\rangle $
with $a$ for different radii based on the equation (\ref{eq:clustering})
with the estimated value of $\delta$ given before. Notably, the clustering
coefficient is predicted to change non-monotonically with the rectangle
side length. Instead, for small values of $a$ the clustering coefficient
is predicted to increase to a maximum value and after it the clustering
decreases linearly. In addition, according to this model, as the connection
radius increases the clustering coefficient is expected to increase
for the same value of $a$. In closing, the largest value of the clustering
is expected for certain specific value of $a$ and a relatively large
connection radius. 

In order to verify these findings we compute the average clustering
coefficient of RRGs with $n=1,500$ and the same connection radii
used in the simulations with the analytical formulas. The results
of the variation of the clustering with the rectangle side length
are illustrated in Fig. \ref{clustering vs a}(b). As can be seen
the clustering increases up to a maximum, whose location depends on
the connection radius, and then decays with the increase of the elongation
of the rectangle. We have not been able to capture the dependence
of $\delta$ with the radius in our previous reasoning, but we very
well captured the \foreignlanguage{american}{behavior} of the clustering
of having a non-monotonic change with $a$. Also, these experiments
show that the increase of the connection radius increases the average
clustering coefficient as predicted by our analytical results. As
can be seen in Fig. \ref{clustering vs a} (b) for $a=1$ and small
radius the average clustering coefficient is $\left\langle C\right\rangle \approx0.61,$
which is very close to the expected value for the 2-dimensional RGG
according to \cite{Dall Christense}. When $a=30$and the radius is
relatively large, the average clustering coefficient is $\left\langle C\right\rangle \approx0.75,$
which coincides with the exact value expected for the one-dimensional
RGG according to \cite{Dall Christense}. Consequently, the RRG generalizes
the values of the clustering coefficient of both, the one- and two-dimensional
RGG, for $a=1$ and $a\rightarrow\infty$, respectively. In addition,
it provides a series of intermediate values of the clustering coefficient
for intermediate values of the side length of the rectangle. 

\begin{figure}[h!]
\begin{centering}
\includegraphics[width=0.4\textwidth]{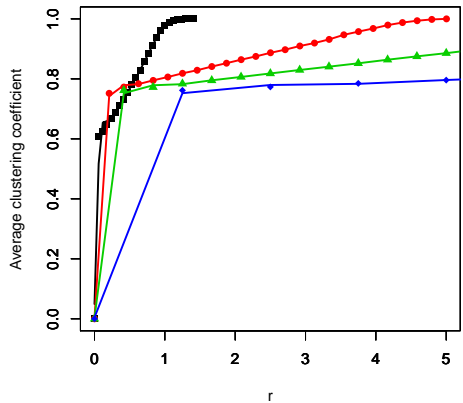}
\par\end{centering}

\protect\protect\protect\caption{(color online) Variation of the average path length with the radius
for RRGs with 1,500 nodes and $a=1,5,10,30$. Every point is the average
of 100 random realizations.}

\label{clustering vs radii}
\end{figure}

We now further explore the relation between the radius $r$ and the
clustering for RRGs with different side lengths. We consider graphs
with $n=1,500$ nodes and $a=1,5,10,30$. As the radius increases
the graph is becoming more and more dense, which makes that the clustering
coefficient is characterized by an abrupt increase at the beginning
of the plot and then a linear increase until the value of $\left\langle C\right\rangle =1$
is reached for the complete graph (see Fig. \ref{clustering vs radii}).

\section*{6 Conclusions and future outlook}

We have introduced here a generalization of the RGG in which we embed
the points into a unit rectangle instead of on a unit square. We consider
a rectangle with sides of lengths $a$ and $b=1/a$, such that as
when $a=1$ we have the particular case of the \textit{classical}
random geometric graph embedded in a unit square. Also, when $a\rightarrow\infty$
we have a very elongated rectangle which resembles a one-dimensional
RGG. We have provided computational and analytical evidence that reaffirm
the fact that the topological properties of the RRG differ significantly
from those of the RGG. In this respect we have obtained analytical
expressions or bounds for the average degree, degree distribution,
connectivity, average path length and the clustering coefficient of
RRGs. In general, these properties depend on the connection radius
as well as on the rectangle side length. Most of the dependencies
found here for these properties in terms of the rectangle side length
are monotonic. The only exception is the clustering coefficient. This
index first increases up to a critical value of $a$ which depends
on the connection radius, and then decays linearly with the increased
elongation of the rectangle.

The introduction of the RRGs opens new possibilities for studying
spatially embedded random graphs. For instance, the study of dynamical
processes taking place on the nodes and edges of these graphs is of
great interest to explore how the shape constraints influence the
dynamics on the RRGs. On the other hand, the analysis of rectangular
proximity graphs, such as the rectangular Gabriel graphs and random
rectangular neighborhood graphs is also interesting for many of the
practical applications of these graphs as mentioned in the Introduction.
The generalization of the RRG model to higher dimensions is also of
both theoretical and practical interest. In closing, the current work
is expected to open new horizons for the study of random spatial graphs
and its applications in physics and beyond.

\section{Acknowledgment}

EE thanks the Royal Society for a Wolfson Research Merit Award. MS
thanks Weir Advanced Research Centre at Strathclyde and EPRSC for
partial financial support of his work. We thank the two anonymous
referees for valuable suggestions that helped to improve this paper.

\end{document}